\documentclass[floats,floatfix,showpacs,amssymb,prd,twocolumn,superscriptaddress,nofootinbib,nolongbibliography,reprint]{revtex4-1}
\usepackage{amssymb,amsmath,verbatim,mathtools,needspace,enumitem,etoolbox,graphicx,physics,microtype,afterpage,tabularx,color,textcmds}

\usepackage{color}
\usepackage{amsmath,amssymb,graphicx}
\usepackage{bm,lipsum}
\usepackage{times}
\usepackage{microtype}
\usepackage[utf8]{inputenc}
\usepackage{booktabs}
\usepackage{subfigure}
\usepackage[normalem]{ulem}
\usepackage[varg]{txfonts}
\usepackage{bm,graphics,graphicx,epsfig,color,ulem}
\definecolor{linkcolor}{rgb}{0.0,0.3,0.5}
\usepackage[unicode, colorlinks=true, linkcolor=linkcolor, citecolor=linkcolor, filecolor=linkcolor,urlcolor=linkcolor, pdfusetitle]{hyperref}
\allowdisplaybreaks
\definecolor{linkcolor}{rgb}{0.0,0.3,0.5}
\newcommand{\ssim}{\mathchar"5218\relax\,}

\def\araa{\rm{ARA\&A}}
\def\apj{\rm{ApJ}}
\def\apjl{\rm{ApJ}}

\def\aapr{\rm{A\&A~Rev.}}

\def\mnras{\rm{MNRAS}}
\def\prd{\rm{PRD}}
\def\prl{\rm{PRL}}
\def\prx{\rm{PRX}}

\def\nat{\rm{Nature}}

\def\cqg{\rm{CQG}}
\def\lrr{\rm{LRR}}

\def\baas{\rm{BAAS}}               %

\newcommand{\penncosmos}{\affiliation{Institute for Gravitation and the Cosmos, Department of Physics, Pennsylvania State University, University Park, PA, 16802, USA}}
\newcommand{\bham}{\affiliation{School of Physics and Astronomy \& Institute for
Gravitational Wave Astronomy, University of Birmingham, Birmingham, B15 2TT, UK}}
\newcommand{\chennai}{\affiliation{Chennai Mathematical Institute, Siruseri, India}}
\newcommand{\jhu}{\affiliation{Department of Physics and Astronomy, Johns Hopkins University, 3400 N. Charles Street, Baltimore, MD, 21218, USA}}
\newcommand{\pennastro}{\affiliation{Department of Astronomy and Astrophysics, Pennsylvania State University, University Park, PA, 16802, USA}}
\newcommand{\cardiff}{\affiliation{School of Physics and Astronomy, Cardiff University, Cardiff, UK, CF24 3AA}}
\newcommand{\flatiron}{\affiliation{Center for Computational Astrophysics, Flatiron Institute, 162 Fifth Ave, New York, NY 10010, USA}}
\newcommand{\suny}{\affiliation{Department of Physics and Astronomy, Stony Brook University, Stony Brook NY 11794, USA}}
\newcommand{\olemiss}{\affiliation{Department of Physics and Astronomy, The University of Mississippi, Oxford MS 38677, USA}}

\begin{document}

\title{Black holes in the low mass gap: Implications for gravitational wave observations}
\author{Anuradha Gupta}\email{agupta1@olemiss.edu}\olemiss\penncosmos

\author{Davide Gerosa}\email{dgerosa@star.sr.bham.ac.uk}\bham

\author{K. G. Arun}\email{kgarun@cmi.ac.in}\chennai \penncosmos 

\author{Emanuele Berti}\email{berti@jhu.edu}\jhu

\author{Will M. Farr}\email{will.farr@stonybrook.edu}\flatiron \suny

\author{B. S. Sathyaprakash}\email{bss25@psu.edu}\penncosmos \pennastro \cardiff

\begin{abstract}
  Binary neutron-star mergers will predominantly produce black-hole
  remnants of mass $\ssim 3$--$4\,M_{\odot}$, thus populating the
  putative \emph{low mass gap} between neutron stars and stellar-mass
  black holes. If these low-mass black holes are in dense
  astrophysical environments, mass segregation could lead to
  ``second-generation'' compact binaries merging within a Hubble time.
  In this paper, we investigate possible 
  signatures of such low-mass compact binary mergers in gravitational-wave observations.
  We show that this unique population of objects, if
present,  will be
  uncovered by the third-generation gravitational-wave detectors, such as
  Cosmic Explorer and Einstein Telescope. Future joint measurements of
  chirp mass ${\cal M}$ and effective spin $\chi_{\rm eff}$ could
  clarify the formation scenario of compact objects in the low mass gap. 
  As a case study, we show that the recent detection of
    GW190425 (along with GW170817) favors a double Gaussian mass model
    for neutron stars, under the assumption that the primary in
    GW190425 is a black hole formed from a previous binary neutron
    star merger.
\end{abstract}


\maketitle

\section{Introduction}
Gravitational wave (GW) observations over the past four years have
brought several exciting discoveries. About half of all black
holes (BHs) discovered by the LIGO and Virgo detectors during their
first and second observing runs had component masses $m_i$ ($i=1,\,2$)
larger than $\sim 30\,M_\odot$, with some of them as massive as
$50\,M_\odot$~\cite{2019PhRvX...9c1040A}. These BH masses are larger
than the BH masses of $\lesssim 25\,M_\odot$ estimated from X-ray
observations~\cite{2017hsn..book.1499C}. Their existence and the
  fact that they would dominate event rates was predicted well before
  their
  discovery~\cite{2010ApJ...714.1217B,2010ApJ...715L.138B,2012ApJ...759...52D,2013ApJ...779...72D,2015ApJ...806..263D}. Eight out
of ten binaries had an effective spin
$\chi_{\rm eff} \equiv (m_1 \chi_{1} \cos\theta_1+ m_2
\chi_{2}\cos\theta_1 )/(m_1+m_2)$ (where $\chi_i$ denotes the Kerr
parameter of each hole and $\theta_i$ is the angle between each spin
and the orbital angular momentum) consistent with zero within the 90\%
credible interval~\cite{2019PhRvX...9c1040A}.  This diversity in
  the mass and spin parameters of LIGO/Virgo binary BHs
hints at a scenario where multiple astrophysical formation channels---including 
isolated binaries \cite{2014LRR....17....3P} and dynamical
interactions \cite{2013LRR....16....4B}---contribute to the observed
population.  Hundreds or thousands of GW events will be required to
assess the relative role of different formation channels 
(see,
e.g.,~\cite{2017PhRvD..95l4046G,2017ApJ...846...82Z,2019MNRAS.488.3810P,2019ApJ...886...25B})
and to probe the BH mass function~\cite{2017PhRvD..95j3010K}.

Theoretical and observational arguments suggest that stellar evolution
may not produce BHs of mass less than
$\ssim
5\,M_\odot$~\cite{1998ApJ...499..367B,2010ApJ...725.1918O,2011ApJ...741..103F}. On
the other hand, neutron stars (NSs) are expected to have a maximum
mass of
$\ssim
3\,M_\odot$~\cite{1974PhRvL..32..324R,2012ApJ...757...55O,2013arXiv1309.6635K,2016arXiv160501665A,2018MNRAS.478.1377A}. The
heaviest NS observed to date has a mass of
$2.01\pm 0.04\,M_\odot$~\cite{2013Sci...340..448A}. There is a recent
claim that PSR J0740+6620 may host a $2.14^{+0.10}_{-0.09}\,M_\odot$ NS, but
systematic uncertainties in this measurement are still a matter of
debate \cite{2020NatAs...4...72C}. 
The lack of observations of compact object in the range
  $\ssim [2,\, 5] M_\odot$ to date hints at the existence of the
  so-called \emph{low mass
  gap}~\cite{1998ApJ...499..367B,2010ApJ...725.1918O,2012ApJ...757...91B},
in contrast with the ``high mass gap'' at $M\gtrsim 50 M_\odot$ due to
pair-instability supernovae~ \cite{2017ApJ...836..244W}.

The general consensus is that the first binary NS merger
GW170817~\cite{2017PhRvL.119p1101A, 2019PhRvX...9c1040A} 
produced a hypermassive
NS~\cite{2017ApJ...850L..19M,2020CQGra..37d5006A} that should
eventually collapse to a BH in the low mass gap (but
see~\cite{2018ApJ...861..114Y, 2019MNRAS.483.1912P} for alternative
possibilities).
If the NS binary progenitors lived in a dense stellar cluster and the BH remnant is retained in the environment, dynamical
interactions and mass segregation could allow it to interact and merge with another compact object. 
The more recent detection of GW190425  \cite{2020arXiv200101761T} hints towards the existence of an unusual 
binary system, if both the binary components are believed to be NSs: 
the total mass of the binary ($3.4^{+0.3}_{-0.1} M_{\odot}$) is significantly larger than the known 
population of galactic double NS binaries~\cite{2019ApJ...876...18F,2019MNRAS.488.5020Z}. 
Moreover, due to the lack of any GW190425 electromagnetic counterpart to date, 
the  GW observation alone can not rule out the possibility of one or 
both binary components being BHs.

The galactic population of double NSs strongly
suggests that a sub-population of detected binaries could well have
been formed in dense clusters (see e.g. Fig. 1 and Sec. 4
of Ref.~\cite{2019ApJ...880L...8A}, where the authors argue that one tenth of the
double NS population in our galaxy has globular cluster
association). Another subpopulation is likely to have been
ejected out of the globular cluster to the galactic field by dynamical
interactions. Hence double NS binaries can exist in dense
environments, and a fraction of these could merge producing low-mass BHs. These, in turn, could pair up with other NSs or 
low-mass BHs and merge within a Hubble time (see~\cite{2020ApJ...888L...3S} for a discussion of the dynamical
interactions which can make this possible), contributing an unknown fraction to the compact binary
population detectable by second- and third-generation GW
detectors. 

Recent globular cluster simulations~\cite{2020ApJ...888L..10Y} predict that binary NSs
dynamically formed in globular clusters should constitute only a tiny 
fraction ($\ssim 10^{-5}$) of the total binary NS mergers, at odds with the observations
mentioned above. More studies are necessary to
determine this fraction, and hence the efficiency of the mechanism
studied in this work. Here we take a model-agnostic stand and
argue that future GW observations will put
this idea to the test, potentially revealing the existence of a new
population of compact binary mergers with one or both components in
the low mass gap. In order to account for these large uncertainties, we
present our results in terms of ``normalized'' detection rates and number
of detections  defined by
${r/f_{\rm dyn}}$ and $N_{\rm det}/f_{\rm dyn}$, respectively, where $r$
and $N_{\rm det}$ are the rate and number of detection of binary NS mergers
and  $f_{\rm dyn}$ denote the fraction of dynamically formed binary NSs
(see Sec.~\ref{sec:assumptions} for details).

In this paper we construct a model to populate the low mass gap with
BHs resulting from a population of merging NSs. 
If binary NSs merge in isolation, the remnant BH that forms in the 
process will never have the opportunity to form a binary again. 
Thus, the mass gap (if it exists) remains intact from the point of view of 
GW observations. Under this assumption, the observation of GW events with 
BH masses in the range $3$--$5M_{\odot}$ would imply that stellar evolution can 
produce BHs in the mass gap. One key point of the present paper is that this conclusion 
could be erroneous if NSs can form in dense clusters.

Following previous work~\cite{2017PhRvD..95l4046G}, we refer to
compact objects born from stellar collapse as ``first generation''
(1g), while ``second generation'' (2g) compact objects are born from
previous mergers.  We will show that 1g+1g, 1g+2g and 2g+2g merger
events in the low mass gap should have rather different chirp mass and
effective spin distributions, that can potentially be distinguished
with third-generation detectors such as Cosmic Explorer
\cite{2019BAAS...51g..35R} and Einstein Telescope
\cite{2010CQGra..27s4002P}.  Similar ideas had previously been
proposed to understand the origin of BHs in the high mass gap, if they
exist in
nature~\cite{2017PhRvD..95l4046G,2019PhRvD.100d1301G,2017ApJ...840L..24F,2018PhRvL.120o1101R}.

The rest of the paper is organized as follows.  In Sec.~\ref{sec:assumptions} we
state our assumptions to model the expected populations of BHs and NSs. In Sec.~\ref{sec:population} we estimate the observational
signatures of binaries containing BHs formed from the merger of binary
NSs. In Sec.~\ref{GW190425} we discuss the implications
of GW190425, if indeed the primary component originated from a 2g merger.
In Sec.~\ref{sec:astro} we conclude the paper and provide directions for future work. 

\section{Building the population}
\label{sec:assumptions}

\begin{figure}[t]
	\centering
	\includegraphics[width=\columnwidth]{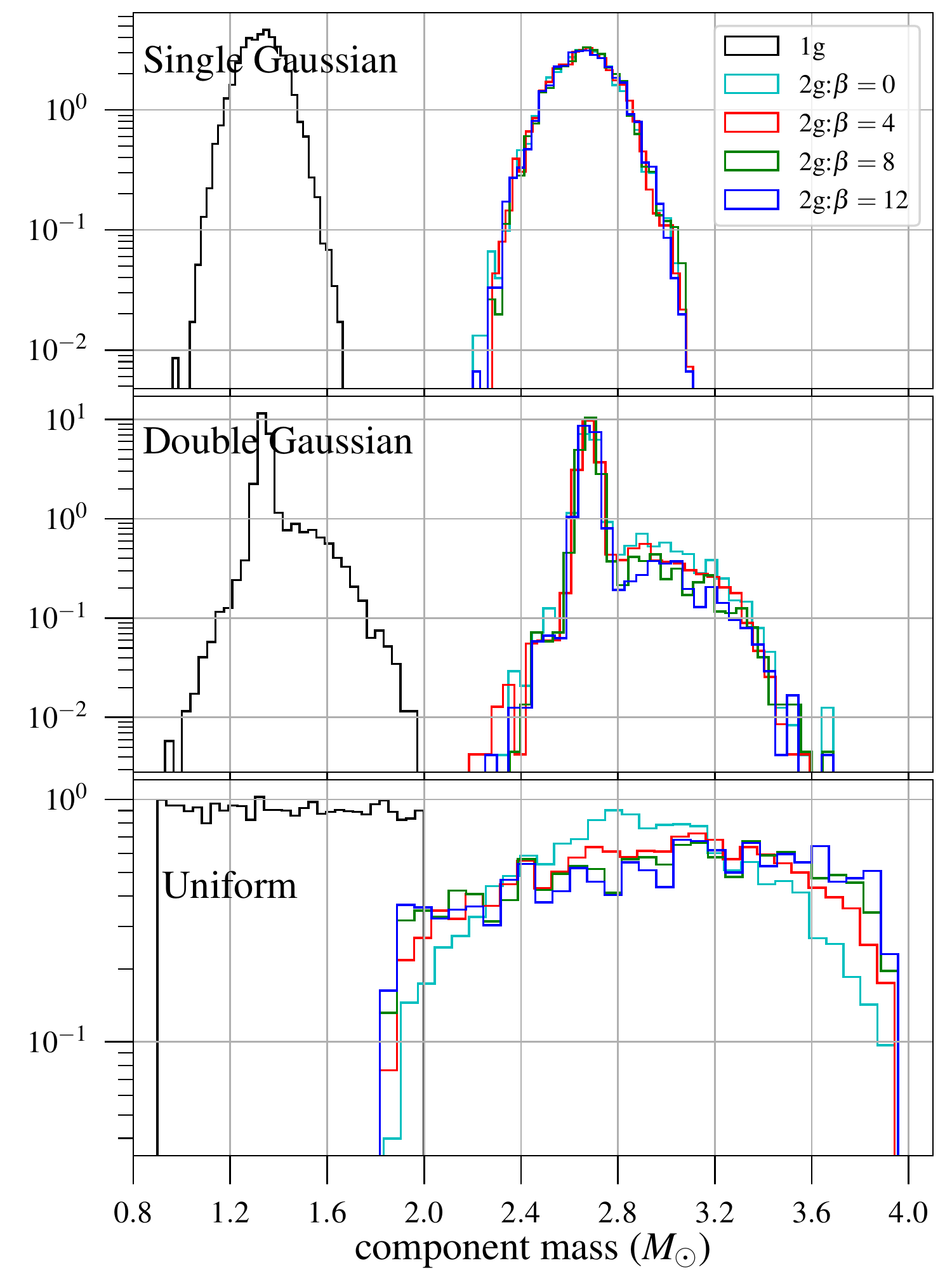}
	\caption{Mass distribution of 1g NSs (black) and 2g BHs (cyan, red, green, blue) for the three mass distributions used in this work and various choices of the pairing-probability spectral index $\beta$.} 
        \label{fig:2gBH}
\end{figure}

The mass distribution of NSs is an active research
topic~\cite{2016ARA&A..54..401O,2016arXiv160501665A,2018MNRAS.478.1377A,2019MNRAS.488.5020Z,2019MNRAS.485.1665K}. 
To bracket uncertainties, we consider three
possibilities~\cite{2016ARA&A..54..401O,2019ApJ...876...18F}: 
\begin{itemize}
\item[(i)] a single Gaussian distribution with mean $\mu= 1.33M_{\odot}$ and standard deviation  $\sigma= 0.09M_{\odot}$;
\item[(ii)] a superposition of two Gaussian distributions with means
  $\mu_i=1.34M_{\odot}, 1.47M_{\odot}$,  standard deviations
  $\sigma_i= 0.02M_{\odot}, 0.15M_{\odot}$, and weights
  $\omega_i=0.68, 0.32$ ($i=1,\,2$), respectively;
\item[(iii)] a uniform distribution in the range
  $[0.9, 2.0]M_{\odot}$.
\end{itemize} We note that model (ii) above is meant to
  reproduce the mass distribution of recycled NSs in double NS
  binaries~\cite{2019ApJ...876...18F}, but
  Ref.~\cite{2019ApJ...876...18F} also reported a uniform distribution
  in the range $[1.14,\, 1.46]M_{\odot}$ for nonrecycled (slowly
  rotating) NSs. Model (iii) in this paper is a more generic uniform
  distribution extending over a broader range.
The resulting mass distributions are shown in Fig.~\ref{fig:2gBH} (black
histograms on the left of the three panels).  We will refer to the
population of NSs drawn from each of these distributions as the
``first generation'' (1g) of compact objects.

Electromagnetic observations indicate that the fastest-spinning
isolated NS has a dimensionless spin magnitude
$\chi\lesssim0.4$~\cite{2006Sci...311.1901H}, while NSs in binaries
are expected to have even smaller spins
$\chi\lesssim
0.04$~\cite{2003Natur.426..531B,2018ApJ...854L..22S}. Therefore, for
simplicity, we will assume our 1g NS population to be nonspinning.

If  formed in dense stellar environments, 1g NSs might  interact with each other,
form binaries, and 
merge. 
We select binaries according to the pairing probability
$p_{\rm pair}(m_1, m_2)\propto (m_2/m_1)^{\beta}$, where
$m_2\leq m_1$. To capture a broad phenomenology, we
vary  the spectral index in the  range $0\leq \beta\leq 12$. This pairing probability is independent of the mass
ratio for $\beta=0$ and favors the formation of comparable-mass
binaries for $\beta>0$. For the case of binary BHs, Refs.~\cite{2019ApJ...882L..24A, 2020ApJ...891L..27F} fitted the observed  GW events  with $p(m_2| m_1)\propto (m_2/m_1)^{\beta}$ and measured $\beta\simeq 7$. 

For a given 1g NS binary characterized by $m_1^{(1{\rm g})}$, $m_2^{(1{\rm g})}$
and selected according to $p_{\rm pair}(m_1^{(1{\rm g})}, m_2^{(1{\rm g})})$, we then
consider their merger product: a ``second-generation'' (2g)
BH. Numerical relativity simulations suggest that the mass ejected in
binary NS mergers is a very small fraction of the total mass of the
system, ranging between $10^{-3}M_{\odot}$ and
$10^{-2}M_{\odot}$~\cite{2019ARNPS..6901918S}.  For simplicity, we neglect the
mass loss and simply estimate the masses of 2g BHs as
$m^{(2{\rm g})} = m^{(1{\rm g})}_1+m^{(1{\rm g})}_2$. The outcome of this procedure is
shown in Fig.~\ref{fig:2gBH}: 2g BHs resulting from the merger of NSs have masses between $\sim 2 M_\odot$ and $\sim 4 M_\odot$
and populate the low mass gap. High (low) values of $\beta$ preferentially
select 1g NSs with comparable (unequal) masses. Their remnants
populate the edges (center) of the 2g mass spectrum.

Binaries containing second generation BHs are expected
to assemble following a sequence of dynamical interactions; therefore
the BH spins in such binaries are expected to be distributed
isotropically. We compute the spin of 2g BHs using fits to numerical-relativity
simulations of BH binaries~\cite{2016ApJ...825L..19H}.\footnote{The final spin
resulting from a NS binary could in principle be quite different from that of a BH binary due to different dependence on mass-ratio, finite-size effects, and nonlinear hydrodynamics contributions.
However, numerical relativity simulations suggests  that the final spin of 
NS binary mergers could be as high as $\sim0.8$ \cite{2013PhRvD..88b1501K,2017PhRvD..95b4029D}, similar to values predicted from BH mergers.
In any case, the use of  BH binary fits for NS binaries does not have
a dramatic impact on our main results.}

\begin{figure*}
	\centering
          \includegraphics[width=\textwidth]{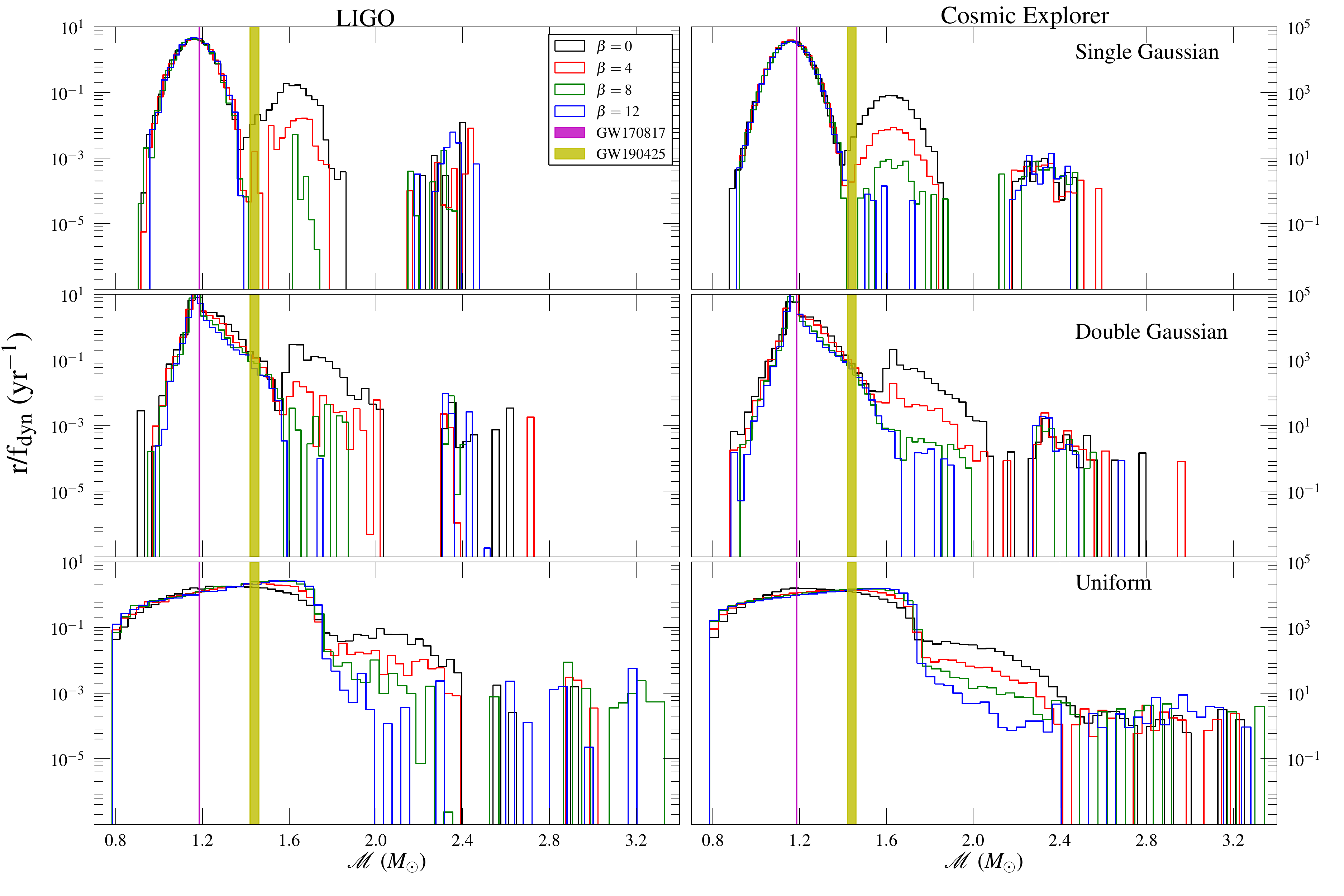}
	\caption{Detection rate $(r/f_{\rm dyn})$ per chirp mass bins for 1g+1g, 1g+2g and 2g+2g
          mergers as observed by LIGO (left) and Cosmic Explorer (right).
          Different colors correspond to different  pairing probabilities
          $p_{\rm pair}\propto (m_2/m_1)^{\beta}$.
          Upper, middle
          and lower panels show results from the three different
          mass distributions for 1g NSs: single-Gaussian,
          double-Gaussian and uniform, respectively. 
          Also shown are the $90\%$ credible bounds on 
          the chirp mass of GW170817 (magenta) and GW190425 (yellow). 
          }
        \label{fig:LIGO_mc}
\end{figure*}

These 2g BHs might interact with the rest of the 1g NSs
in the population or with other 2g BHs,
form binaries, and possibly produce GW events.
If BHs heavier than $5\,M_\odot$ formed by stellar evolution reside in
the same cluster, they too could pair up with 2g BHs, and produce merger
events with a more extreme mass ratio. 
In fact, due to the higher
rate of BH mergers compared to NS mergers in dense stellar 
environments \cite{2016PhRvD..93h4029R,2020ApJ...888L..10Y}, it is likely that most
low mass gap binaries will come from the merger of 2g BHs with
massive BHs outside the gap ($\gtrsim 5M_\odot$).
However, we do not consider this possibility in the present work.
In our model, the initial population is that of NS binaries and they
are the ones that produce BHs in the mass gap. The existence of
heavier BHs in the same environment could alter the distribution of BH
masses and spins to be discussed below, and we plan to investigate
this problem in the future.

Let us make a rough estimate of the abundance of 2g BHs by assuming
that they are produced continuously since the formation of the first
galaxies.  The Milky Way has $\ssim 10^8$
NSs~\cite{1995A&ARv...6..225H}.  The detection of GW170817 and GW190425 has established the rate of
binary NS mergers to be $\ssim 10^3\, \rm Gpc^{-3} \rm yr^{-1}$, which
translates to a merger rate in a Milky Way Equivalent Galaxy of
$10^{-4}\,\rm yr^{-1}$~\cite{2010CQGra..27q3001A}. Within the age of
the Universe $\ssim 10\,\rm Gyr$, we expect that such a galaxy would
have witnessed as many as $10^6$ binary NS mergers, leading to the same
number of 2g BHs. Thus, the abundance ratio of 2g BHs to NSs in the
Universe could be assumed to be $\kappa=0.01$.

This yields a mixture population
\begin{equation}
p(m) = (1-\kappa)\, p(m^{(1{\rm g})})+\kappa \, p(m^{(2{\rm g})})\,.
\end{equation}
From this distribution, we extract two masses $m_1$ and $m_2$ according to
$p_{\rm pair}$ and consider their GW emission. This constructs the probability 
distribution $p(m_1, m_2)$ and leads to three
populations: 1g+1g (where both companions are NSs), 2g+2g (where both
companions are BHs), and 1g+2g (where a NS pairs with a BH).

We distribute merger redshifts uniformly in comoving volume $V_c$ and
source-frame time, i.e.  $p(z) \propto (dV_c/dz)/(1+z)$, up to some
horizon redshift $z_{\rm H}$.  We estimate GW detectability using a
standard single-detector semi-analytic
approximation~\cite{1993PhRvD..47.2198F,2015ApJ...806..263D} 
with a signal-to-noise ratio (SNR) threshold of 8 and the waveform
model of \cite{2016PhRvD..93d4007K}. This defines a 
detection probability $p_{\rm det}(m_1,m_2,z)$ averaged over polarization,
inclination, and sky location.  We neglect spins, because they have a
small effect on the detection rate~\cite{2018PhRvD..98h4036G}. We
consider noise curves for advanced LIGO at design
sensitivity~\cite{2016LRR....19....1A} and Cosmic Explorer in the
wide-band configuration~\cite{2017CQGra..34d4001A}.  The horizon
redshift $z_{\rm H}$ is chosen such that $p_{\rm det}=0$ for
$z>z_{\rm H}$. In particular, we set $z_{\rm H} = 0.3$ (4) for LIGO
(Cosmic Explorer).

The expected merger rate is given by
\begin{align}
r\!= \!\!\int\!\!  p(m_1,m_2)
{\cal R}(z) \frac{dV_c}{dz} \frac{1}{1+z}
 p_{\rm det}(m_1, m_2, z)  {\rm d}m_1  {\rm d} m_2{\rm d} z\,,
\label{rateintegral}
\end{align}
where ${\cal R}(z)$ is the intrinsic merger rate. If ${\cal R}_{\rm NS}$ is the total NS-NS merger rate,  
only the fraction $f_{\rm dyn}$ coming from dynamical channels is relevant to the formation 
of 2g BHs, i.e., ${\cal R}=f_{\rm dyn} {\cal R}_{\rm NS}$. At
  present, there is no clear consensus on the value of $f_{\rm dyn}$: see e.g.
\cite{2019ApJ...880L...8A} and \cite{2020ApJ...888L..10Y}.
Therefore we quote ``normalized'' merger rates $r/f_{\rm dyn}$, and we assume 
${\cal R}_{\rm NS}$ to be $1000$  Gpc$^{-3}$yr$^{-1}$ \cite{2020arXiv200101761T}. 
Our results can be easily rescaled when future events will 
better constrain these values.
In practice, we approximate Eq.~(\ref{rateintegral}) with a Monte-Carlo sum
\begin{equation}
\frac{r}{f_{\rm dyn}} \approx  {\cal R_{\rm NS}}  \Bigg(\!\!\int_{0}^{z_{\rm H}}\!\! \frac{dV_c}{dz} \frac{1}{1+z} dz \!\Bigg) \frac{1}{N} \sum_{i=1}^{N}\, p_{\rm det}(m_1^{i}, m_2^{i}, z^{i}) ,
\end{equation}
where $N$ is the total number of simulated binaries.
The total number of observations, scaled with $f_{\rm dyn}$, is then given by $N_{\rm det}/f_{\rm dyn}  = (r/f_{\rm dyn}) \times T_{\rm obs}$,
where $T_{\rm obs}$ is the duration of the observing run(s).

\section{Filling the mass gap}
\label{sec:population}

Figure~\ref{fig:LIGO_mc}
shows histograms of the detection rate as a function of the chirp mass ${\cal M}=(m_1m_2)^{3/5}/(m_1 + m_2)^{1/5}$.
The predicted distribution presents three distinct peaks at low,
moderate and high values of ${\cal
  M}$: in these regimes the merger rate is dominated by 1g+1g, 1g+2g
and 2g+2g mergers, respectively. The ratio between the height of the
peaks is $\sim \kappa$ and $\sim
\kappa^2$, as a consequence of the rate argument presented above.
Among the three populations, hybrid (1g+2g) mergers present the
strongest dependence on
$\beta$.  These mergers are characterized by mass ratios
$\ssim0.5$, which are suppressed for steep pairing probability
functions. %

Clearly, if NS masses are distributed with a single-peak Gaussian (which could be confirmed with future observations) then even 1g-2g and 2g-2g mergers continue to leave a gap in chirp mass between $\sim 1.8\,M_\odot$ and $\sim 2.2\,M_\odot.$ This would be absent if massive stars are able to leave a remnant in the mass gap. Since we will be able to measure the NS mass distribution very accurately with future detections, this is a firm prediction about the existence of the mass gap that could be tested with third-generation GW detectors.

\begin{figure}[t]
	\centering
	\vspace{0.5cm}
	\includegraphics[width=\columnwidth]{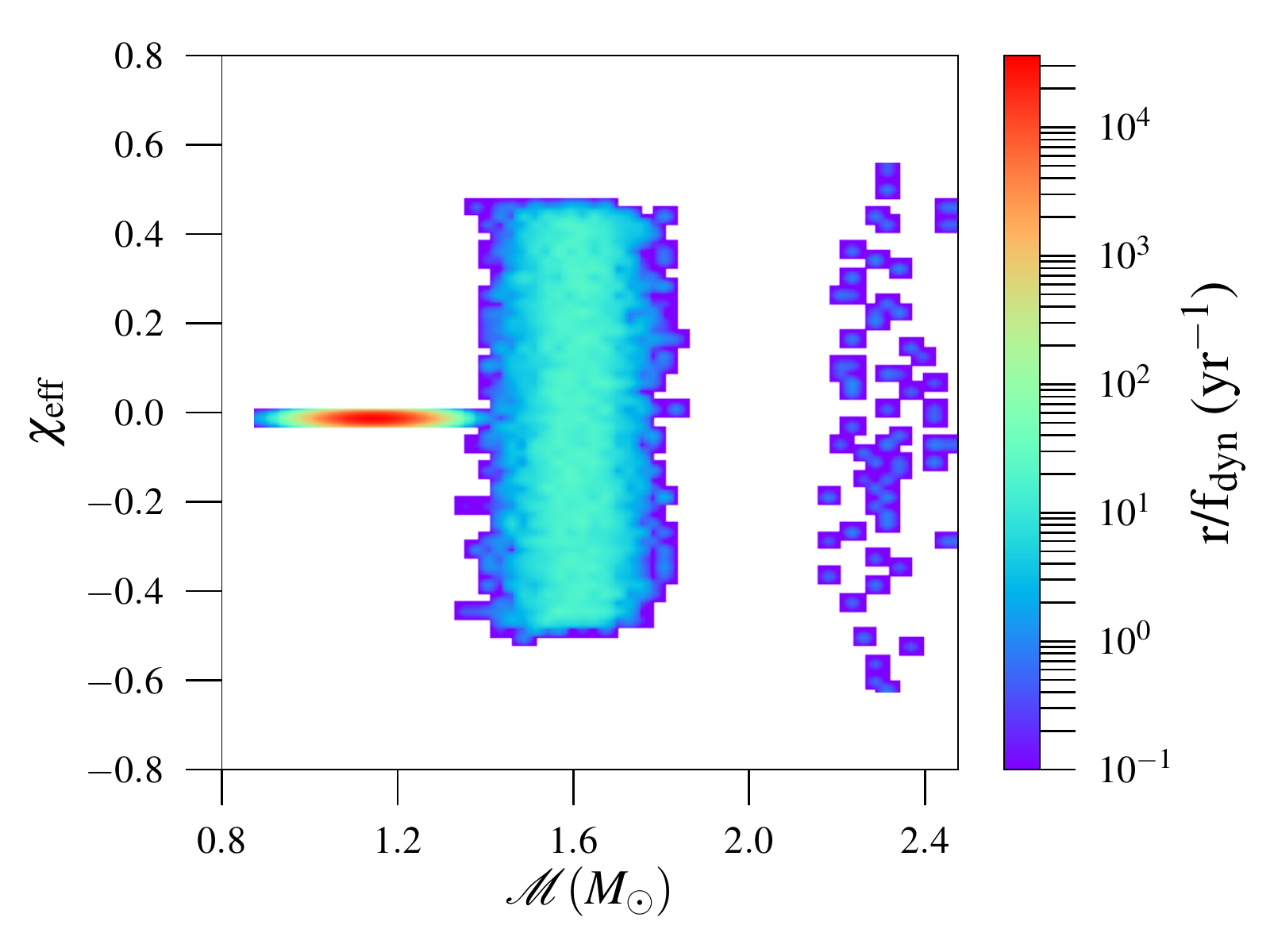}
	\caption{Joint chirp-mass effective-spin distribution as observed
          by Cosmic Explorer. The 1g NS mass distribution is modeled by a single
          Gaussian and we assume $\beta=0$. The color bar indicates the detection rate $(r/f_{\rm dyn})$ per bin.}
        \label{fig:mc_chieff}
 \end{figure}

Figure~\ref{fig:mc_chieff} shows the joint distribution of chirp mass and effective spin
 observed by Cosmic Explorer
assuming the single Gaussian mass distribution and $\beta =0$; results
are qualitatively similar under other assumptions.

Again, events separate into three distinct regions, corresponding to  1g+1g, 1g+2g, and 2g+2g.
At low chirp masses, the event rate is dominated by 1g+1g NS mergers,
which are slowly rotating. The effective spin is thus expected to be
very small (exactly zero in our simplified model).  At moderate chirp
masses, the rate is dominated by 1g+2g mergers. In these BH/NS
mergers, the BH is the result of a nonspinning comparable-mass merger,
and therefore has $\chi_1\ssim 0.7$ \cite{2007PhRvD..76f4034B}. The
NS, on the other hand, has $\chi_2\ssim 0$. Since we neglect mass loss
and NSs have a relatively narrow mass distribution, these events have
$m_1\ssim 2 m_2$. The largest (smallest) effective spins these events
can have correspond to $\theta_1=0$ ($\pi$), which implies
$|\chi_{\rm eff}|\gtrsim 2 \times 0.7/(1+2)\ssim 0.45$, as shown
in Fig.~\ref{fig:mc_chieff}.  Events with ${\cal M}\gtrsim 2M_\odot$
are 2g+2g BH mergers. In this case, $\chi_1\ssim \chi_2\ssim 0.7$ and
$m_1\ssim m_2$. The effective spin is bound by
$|\chi_{\rm eff}| \lesssim 0.7$.

\begin{table}
\centering
\begin{tabular}{c|cccccc}
&   \multicolumn{3}{c}{LIGO}  &  \multicolumn{3}{c}{Cosmic Explorer} \\ 
\hline
$\beta$  & 1g+1g & 1g+2g & 2g+2g & 1g+1g & 1g+2g & 2g+2g\\ 
\hline
  \multicolumn{7}{c}{Single-peak Gaussian mass distribution } \\ 
  \hline
  0 & 24 & 1.06 & 0.01 & $1.9\times10^{5}$ & 5443 & 45 \\  

  2 & 24 &  0.30 & 0.02  &  $1.9\times10^{5}$ & 1695  & 39 \\ 

  4 & 25 & 0.10  & 0.01 & $1.9\times10^{5}$  & 516 &  43 \\

  6 & 25 &  0.02 & 0.02  & $1.9\times10^{5}$ & 149 & 49 \\ 

  8 & 25  & $5\times10^{-3}$ & $2\times10^{-3}$ & $1.9\times10^{5}$ & 47 & 43 \\ 

  10 & 25 & $4\times10^{-3}$  & 0.04 & $1.9\times10^{5}$  & 21 & 39 \\ 

  12 & 25 & 0 & 0.01 & $2.0\times10^{5}$  & 3 & 50\\ 
\hline
  \multicolumn{7}{c}{Double-peak Gaussian mass distribution } \\ 
  \hline
  0 & 26 & 1.19 & 0.01 & $2.0\times10^{5}$ & 5804 & 43 \\  

  2 & 26 &  0.42 & $8\times10^{-3}$  &  $2.1\times10^{5}$ & 1675  & 38 \\ 

  4 & 26 & 0.08  & $5\times10^{-3}$ & $2.1\times10^{5}$  & 522 &  49 \\

  6 & 26 &  0.04 & 0.02  & $2.1s\times10^{5}$ & 153 & 51 \\ 

  8 & 26  & 0.01 & $9\times10^{-3}$ & $2.0\times10^{5}$ & 41 & 32 \\ 

  10 & 26 & $3\times10^{-3}$  & 0.01 & $2.0\times10^{5}$  & 15 & 47 \\ 

  12 & 26 & $9.9\times10^{-5}$ & 0.02 & $2.0\times10^{5}$  & 8 & 54\\ 
\hline
  \multicolumn{7}{c}{Uniform mass distribution } \\ 
  \hline
  0 & 29 & 1.32 & 0.01 & $2.1\times10^{5}$ & 6042 & 39 \\  

  2 & 30 &  0.64 & 0.03  &  $2.2\times10^{5}$ & 2748  & 52 \\ 

  4 & 31 & 0.31  & $6.9\times10^{-3}$ & $2.2\times10^{5}$  & 1482 &  35 \\

  6 & 32 &  0.19 & $4\times10^{-3}$  & $2.2\times10^{5}$ & 959 & 40 \\ 

  8 & 32  & 0.15 & 0.02 & $2.2\times10^{5}$ & 673 & 52 \\ 

  10 & 33 & 0.17  & 0.03 & $2.2\times10^{5}$  & 625 & 48 \\ 

  12 & 33 & 0.13 & 0.01 & $2.2\times10^{5}$  & 576 & 51\\ 
\hline
\end{tabular}
\caption{Expected number of detections $N_{\rm obs}/f_{\rm dyn}$ from one year of observation of LIGO at design sensitivity and Cosmic Explorer.}
\label{Ndet}
\end{table}

Table~\ref{Ndet} shows the expected number of observations, assuming one
year of data from either advanced LIGO at design sensitivity or Cosmic
Explorer.  With
second-generation interferometers, the 
expected number of observations for this population of BHs in the low
mass gap is extremely small. Third-generation
detectors will be necessary to unveil these systems, thus
adding yet another item to their already vast science case~
\cite{2019BAAS...51g..35R,2010CQGra..27s4002P}.
In a few  years of operation, Cosmic Explorer might deliver between
$\ssim 1$ and $\ssim100$~BHs in the low mass gap if $f_{\rm dyn} \sim 0.01$. A few events should 
still be visible even if the dynamical contribution to the NS merger 
rate is smaller than $0.01$. 

\section{GW190425: A case study}
\label{GW190425}

LIGO and Virgo have announced the detection of a new NS binary,
GW190425 \cite{2020arXiv200101761T}, during their third observing run. This
is the second event, after GW170817, that is believed to contain at
least one NS component (or possibly two). This inference is based
entirely on the measured masses of the binary components,
$m_1 \in [{1.61}, {2.52}]$ and $m_2\in [{1.12, 1.68}].$ As
acknowledged in \cite{2020arXiv200101761T}, GW observation alone cannot
rule out the possibility that this is a NS-BH or a BH-BH binary. A
nonzero value for the tidal deformability $\tilde \Lambda$ would be a
signature of the presence of at least one NS in the system, while
$\tilde\Lambda$ consistent with zero would imply that the system could
be a BH binary. The signal-to-noise ratio for GW190425 is not large
enough to infer that $\tilde\Lambda$ is nonzero, thus not ruling out the
possibility that one (or both) of the components in GW190425 is a BH.

What are the astrophysical implications if the primary component is a BH? In
particular, can we say anything about the merger rate of such a BH
with NSs?  As discussed before, it is likely that no compact objects
are produced by stellar evolution in the mass range
$[2.2,\,5]\,M_\odot,$ but NS mergers would definitely produce BHs in
the mass gap.  Furthermore, if binary NS mergers occur in globular
clusters, then the resulting BHs could merge with other NSs in the
cluster.  Under the assumption that GW190425 is the result of such a
merger, we can estimate the rate of these second-generation mergers
relative to the binary NS merger rate.

\begin{figure}[]
\centering
\includegraphics[width=0.5\textwidth]{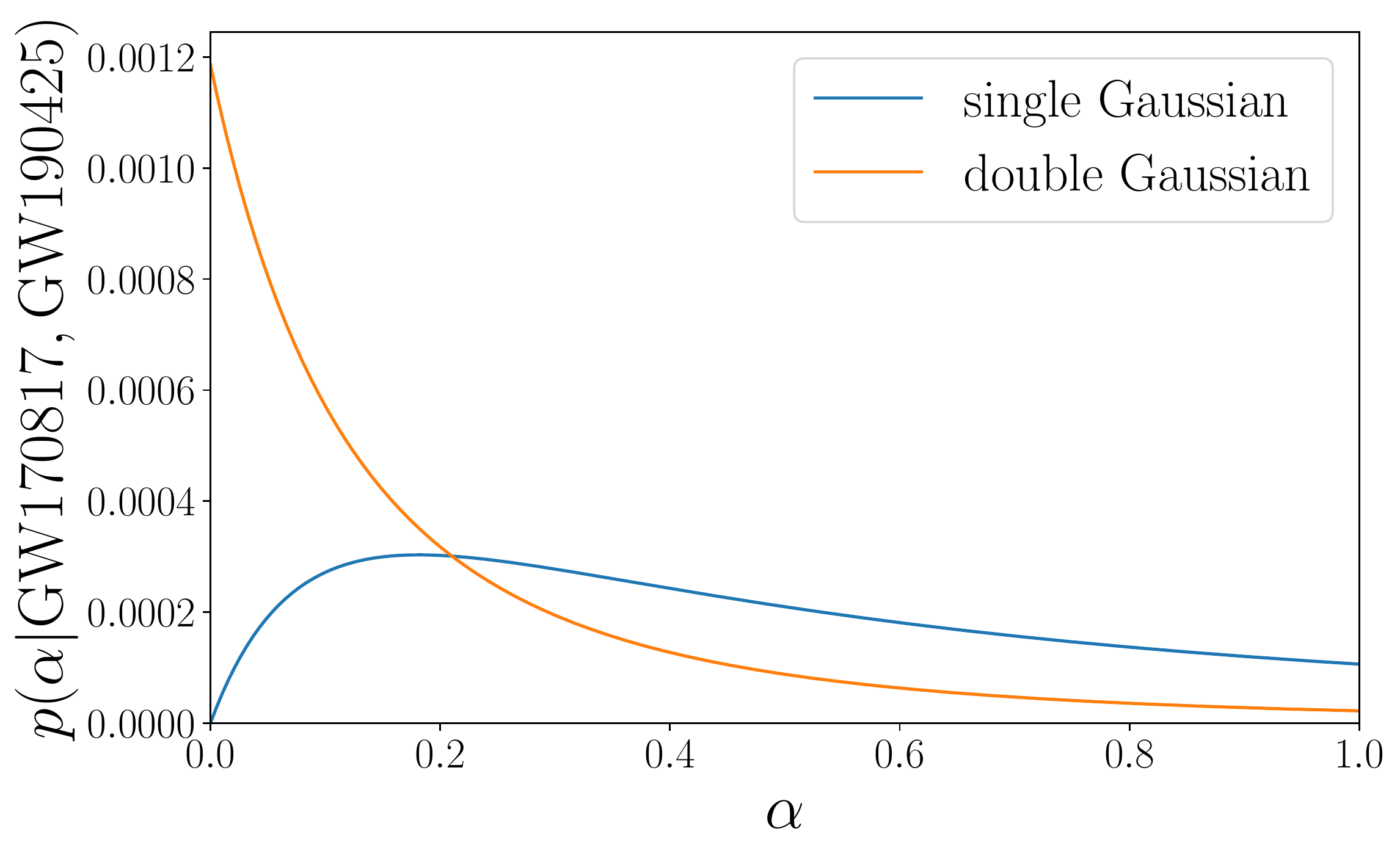}
\caption{The posterior probability of $\alpha$ computed using Eq.~(\ref{eq:p_alpha2}).}
\label{fig:alpha}
\end{figure}
In the absence of any other process (e.g.,
stellar evolution or primordial BHs) contributing to the BH population in the
mass gap, an observed merger belongs to one of two classes: (a) a NS-NS merger,
that we call a class 1g merger, or (b) a NS-BH merger (let us call it a ``class 1.5g'' merger) where
the BH is the result of a previous NS-NS merger. If $\alpha$ denotes the rate of 1.5g mergers
relative to 1g mergers, then $R_{1.5} = \alpha\, R_1$. 
Assuming that the sensitive volumes for the
two mergers are $V_1$ and $V_{1.5}$, respectively, the Poisson likelihood for these detections is~\cite{2015PhRvD..91b3005F,2018PhRvD..98h3017T,2019MNRAS.486.1086M}
\begin{equation}
p\left( N_1, N_{1.5} \mid R_1, \alpha \right) = \alpha^{N_{1.5}} R_1^{N_1 + N_{1.5}}
 e^{- R_1 \left( V_1 + \alpha V_{1.5} \right)},
\end{equation}
where $N_1$ and $N_{1.5}$ are the number of 1g and 1.5g mergers, respectively.
Integrating out $R_1$ after applying a prior $p\left( R_1 \right) \propto R_1^{-\gamma}$ (e.g.,
$\gamma = 0$ for a flat prior, $\gamma = 1/2$ for Jeffreys prior, $\gamma = 1$ for a flat-in-log
prior), we find
\begin{equation}
\label{eq:p_alpha}
p\left( N_1, N_{1.5} \mid \alpha \right) \propto \frac{\alpha^{N_{1.5}}}{\left( V_1 + \alpha V_{1.5} \right)^{N_1 + N_{1.5} - \gamma + 1}}\,.
\end{equation}
If we make the approximation that the volumes scale in the usual way with chirp mass, then
Eq.~(\ref{eq:p_alpha}) becomes 
\begin{equation}
\label{eq:p_alpha2}
p\left( N_1, N_{1.5} \mid \alpha \right) \propto \frac{\alpha^{N_{1.5}}}{\left[ 1 + \alpha \left( \frac{\mathcal{M}_{1.5}}{\mathcal{M}_1} \right)^{15/6}\right]^{N_1 + N_{1.5}- \gamma + 1}}\,,
\end{equation}
where $\mathcal{M}_1$ and $\mathcal{M}_{1.5}$ are characteristic chirp masses for 1g and 1.5g
populations, respectively.

Based on the chirp mass measurement of GW170817 and GW190425, and referring to Fig.~\ref{fig:LIGO_mc},
it is likely that $N_1=1$ and $N_{1.5}=1$ for single Gaussian mass-model 
(GW170817 belonging to 1g class and GW190425 belonging to 1.5g class), and
$N_1=2$ and $N_{1.5}=0$ for double Gaussian mass-model (both GW170817 and GW190425 belonging to 1g class) \cite{2019ApJ...876...18F}. 
Assuming a flat prior on $R_1$ 
(i.e., $\gamma=0$) and taking $\mathcal{M}_{1.5}/\mathcal{M}_{1}=1.5$, we compute
the posterior probability for $\alpha$ for the two mass models in Fig.~\ref{fig:alpha}.

The posterior probability for $\alpha$ gives a 90\% credible interval
for $\alpha$ of $[0.08,\,0.91]$ for the single Gaussian mass model,
whereas $\alpha>8.2\times10^{-5}$ for the double Gaussian models.  The
current binary evolutionary models predict very small $\alpha$ values \cite{2020ApJ...888L..10Y},
consistent with the value inferred above using the double Gaussian
mass model.  The single Gaussian model, on the other hand, provides
relatively large $\alpha$. This implies that the two NS events so far
(GW170817 and GW190425) favor a double Gaussian mass model over the
single Gaussian mass model, under the assumptions that (i) BHs in the
mass gap are formed only via compact binary mergers, and (ii) the
primary of GW190425 is a BH formed through a NS-NS merger. Future
observations would either strengthen or weaken this claim.

\section{Concluding remarks and Future Work}
\label{sec:astro}

Astrophysical considerations suggest the possible existence of a mass gap 
between the heaviest NSs and the lightest stellar-origin
BHs~\cite{1998ApJ...499..367B,2010ApJ...725.1918O}.
The gap could well be just a selection
effect~\cite{2011ApJ...741..103F}, so it is important to verify
whether BHs populating the mass gap exist in nature. GW observations
will present orthogonal selection effects compared to electromagnetic
probes, thus offering a promising opportunity to settle this issue.  As
the number of GW detections increases, we will be able to determine
whether the mass gap is populated, and to set constraints on the
astrophysical mechanisms that populates it~\cite{2015ApJ...807L..24L,2015MNRAS.450L..85M}.

Understanding the existence of compact objects in the mass gap has
important astrophysical implications. Stellar collapse can only
produce BHs with masses $M\lesssim 5M_\odot$ if the explosions are
driven by instabilities that develop over time scales
$\gtrsim 200$~ms~\cite{2012ApJ...757...91B}: if these instabilities
develop on shorter time scales, the predicted mass spectrum has a gap.

Several arguments indicate that the first binary NS merger
GW170817 must have produced a hypermassive
NS~\cite{2017ApJ...850L..19M,2020CQGra..37d5006A}, that should
eventually collapse to a BH in the low mass gap.  The total
  mass of the GW190425 binary is significantly larger than the mass of
  galactic double NS binaries, and we can not rule out the possibility
  of one or both binary components being BHs. These two observed
  events and simple rate estimates suggest that the ratio of NS-NS
merger remnants to NSs in a Milky Way Equivalent Galaxy should be
$\kappa\ssim 0.01$.  This implies the existence of a population of
low-mass BHs in merging compact binaries, which can be probed with
third-generation GW detectors.

The inverse problem is also intriguing. Measuring the relative
abundance of NS mergers and low mass gap BH mergers will allow us to
infer the typical number of NS mergers occurring in a galaxy during
its cosmic lifetime.

There is one caveat in our models: we assume that all merger remnants
are retained inside the cluster and remain available to form 2g
objects. Both natal and merger kicks might decrease the available
number of low-mass BHs in clusters. Including this effect in future
work might provide a handle to constrain the escape speed of dense
stellar clusters with GW data~\cite{2019PhRvD.100d1301G}.

Some events
(e.g. GW151226 and GW170608; \cite{2019PhRvX...9c1040A}) already
hint at a non-negligible probability that some BHs may be in the low
mass gap. At the present sensitivity, however, those posterior tails
strongly depends on the assumed prior~\cite{2017PhRvL.119y1103V}.
 We plan to explore the astrophysical implications of this 
population of BH binaries in future work.

\section*{Acknowledgments}
We thank Simon Stevenson and Nathan Johnson-McDaniel for carefully
reading the manuscript and providing useful comments and Surabhi 
Sachdev for discussion and comments. A.G. and B.S.S. are supported 
in part by NSF grants PHY-1836779, AST-1716394 and AST-1708146. 
D.G. is supported by Leverhulme Trust Grant No. RPG-2019-350.
 K.~G.~A. is partially supported by the Swarnajayanti
Fellowship Grant No. DST/SJF/PSA-01/2017-18 and a grant from Infosys
Foundation.  E.B. is supported by NSF grant No. PHY-1841464, NSF grant
No. AST-1841358, NSF-XSEDE grant No. PHY-090003, NASA ATP grant
No. 17-ATP17-0225, and NASA ATP grant No.  19-ATP19-0051.  This
research was supported in part by the NSF under Grant No. NSF
PHY-1748958.  E.B. acknowledges support from the Amaldi Research
Center, funded by the MIUR program ``Dipartimento di
Eccellenza''~(CUP: B81I18001170001), and thanks the physics department
at the University of Rome ``Sapienza'' for hospitality during the
completion of this work.  This project has received funding from the
EU H2020 research and innovation programme under the Marie
Sklodowska-Curie grant agreement No 690904.  The authors acknowledge
networking support by the COST Action CA16104 ``GWverse''.
Computational work was performed on the University of Birmingham's
BlueBEAR cluster, the Athena cluster at HPC Midlands+ funded by EPSRC Grant No. EP/P020232/1, the Maryland Advanced Research Computing Center
(MARCC), and the IUCAA LDG cluster Sarathi.  The evaluation of
$p_{\rm det}$ was performed with the \textsc{gwdet} code available at
\href{https://github.com/dgerosa/gwdet}{github.com/dgerosa/gwdet}.
This document has LIGO preprint number {\tt P1900271}.

\bibliography{low_massgap}
\end{document}